\documentclass[english,aps,pre,twocolumn,amsmath,amssymb,amsfonts]{revtex4}
\usepackage[T1]{fontenc}
\usepackage[latin9]{inputenc}
\usepackage{array}
\usepackage{verbatim}
\usepackage{bm}
\usepackage{amsbsy}
\usepackage{amssymb}
\usepackage{graphicx}
\usepackage{esint}

\makeatletter

\providecommand{\tabularnewline}{\\}

\@ifundefined{textcolor}{}
{%
 \definecolor{BLACK}{gray}{0}
 \definecolor{WHITE}{gray}{1}
 \definecolor{RED}{rgb}{1,0,0}
 \definecolor{GREEN}{rgb}{0,1,0}
 \definecolor{BLUE}{rgb}{0,0,1}
 \definecolor{CYAN}{cmyk}{1,0,0,0}
 \definecolor{MAGENTA}{cmyk}{0,1,0,0}
 \definecolor{YELLOW}{cmyk}{0,0,1,0}
}

\usepackage{pslatex}

\makeatother

\usepackage{babel}
\begin{document}

\title{Microstructure identification via detrended fluctuation analysis
of ultrasound signals}

\author{Paulo G. Normando}

\author{Romão S. Nascimento}

\author{Elineudo P. Moura}

\email{elineudo@ufc.br}

\address{Departamento de Engenharia Metalúrgica e de Materiais, Universidade
Federal do Ceará, 60455-760, Fortaleza, CE, Brazil}

\author{André P. Vieira}

\email{apvieira@if.usp.br}

\address{Instituto de Física, Universidade de São Paulo, Caixa Postal 66318,
05314-970, São Paulo, SP, Brazil}

\date{\today}
\begin{abstract}
We describe an algorithm for simulating ultrasound propagation in
random one-dimensional media, mimicking different microstructures
by choosing physical properties such as domain sizes and mass densities
from probability distributions. By combining a detrended fluctuation
analysis (DFA) of the simulated ultrasound signals with tools from
the pattern-recognition literature, we build a Gaussian classifier
which is able to associate each ultrasound signal with its corresponding
microstructure with a very high success rate. Furthermore, we also
show that DFA data can be used to train a multilayer perceptron which
estimates numerical values of physical properties associated with
distinct microstructures. 
\end{abstract}
\maketitle

\section{Introduction}

Much attention has been given to the problem of wave propagation in
random media by the condensed-matter physics community, especially
in the context of Anderson localization and its analogs \cite{anderson1958,hodges1982,baluni1985,gavish2005,bahraminasab2007}.
A hallmark of these phenomena is the fact that randomness induces
wave attenuation by energy confinement, even in cases where dissipation
can be neglected. The interplay of energy confinement and randomness
gives rise to noisy but correlated signals, for instance, in the case
of acoustic pulses propagating in inhomogeneous media. 

It has long been known that the correlations in a time series hide
relevant information on its generating dynamics. In a pioneering paper,
Hurst \cite{hurst51} introduced the rescaled-range analysis of a
time series, which measures the power-law growth of properly rescaled
fluctuations in the series as one looks at larger and larger time
scales $\tau$. The associated Hurst exponent $H$ governing the growth
of such fluctuations is able to gauge memory effects on the underlying
dynamics, offering insight into its character. The presence of long-term
memory leads to a value of the exponent $H$ which deviates from the
uncorrelated random-walk value $H=1/2$, persistent (antipersistent)
behavior of the time series yielding $H>1/2$ ($H<1/2$). Additionally,
a crossover at a time scale $\tau_{\times}$ between two regimes characterized
by different Hurst exponents may reveal the existence of competing
ingredients in the dynamics, and in principle provides a signature
of the associated system. This forms the base of methods designed
to characterize such systems \cite{matos04}, if one is able to obtain
reliable estimates of the Hurst exponents. This is in general a difficult
task, as even local trends superposed on the noisy signal may affect
the rescaled-range analysis, obscuring the value of $H$. A related
exponent, $\alpha$, defined through detrended fluctuation analysis
(DFA) \cite{peng94}, can be used instead.

Actually, the characteristics of exponents and crossovers observed
in the DFA curves associated with various types of data series have
been extensively used to distinguish between the systems producing
such series. Examples include coding versus noncoding DNA regions
\cite{peng94,ossadnik1994}, healthy versus diseased subjects as regards
cardiac \cite{goldberger2002}, neurological \cite{hausdorff2000,hwa2002}
and respiratory function \cite{peng2002}, and ocean versus land regions
as regards temperature variations \cite{fraedrich2003}. However,
there are many instances in which these characteristics are not clearcut
and the DFA curves exhibit a more complex dependence on the time scale.
In such cases, it has been shown that pattern recognition tools \cite{webb02}
can help the identification of relevant features, greatly improving
the success of classification tasks \cite{vieira08,vieira2009qnde,moura2009,vieira2010}.

In the present work, we investigate the possibility of extracting
information on the nature of inhomogeneities by analyzing fluctuations
in time series associated with ultrasound propagation in random media.
A hint that this possibility is real was provided by the fact that
the crossover features of DFA and Hurst analysis curves from backscattered
ultrasound signals revealed signatures of the microstructure of cast-iron
samples \cite{matos04}. Here, in order to perform a systematic study,
we resort to simulating the propagation of ultrasound pulses in one-dimensional
media with distinct microstructures, defined by probability distributions
of physical properties such as domain size, density and sound velocity.
Although this choice of geometry cannot allow for the full phenomenology
of sound propagation (such as mode conversion from transverse to longitudinal
sound waves), it makes it possible to generate large quantities of
simulated data, which are important in order to assess the generalizability
of our reported results. Moreover, it approximately describes normal
propagation of sound waves in layered media. 

The paper is organized as follows. In Sec. \ref{sec:ultrasound} we
present the artificial microstructures we produced, as well as a sketch
of the simulation technique; a detailed description is relegated to
Appendix \ref{sec:appendix}. In Sec. \ref{sec:dfa} we describe the
method of detrended fluctuation analysis and its results when applied
to our simulated signals. Then, in Sec. \ref{sec:gaussiandiscriminants},
we report on an automated classifier which is able to associate, with
a very high success rate, the DFA curves with the corresponding microstructure.
Furthermore, we show in Sec. \ref{sec:neuralnet} that the DFA curves
can be used to train a neural network which predicts numerical values
of physical properties associated with different microstructures.
We close the paper by presenting a summary in Sec. \ref{sec:conclusions}.

\section{Simulating ultrasound propagation}

\label{sec:ultrasound}We are interested in studying ultrasound propagation
along a one-dimensional medium of width $W$, with a pulse generated
in a transducer located at one end of the system. Since the medium
consists of many different domains, with possibly different physical
properties (density and sound velocity), in general the pulse will
be scattered as it propagates towards the opposite end, where it will
be reflected. Information on the microstructure is in principle hidden
in the scattered signal, which is registered in the transducer as
it arrives.

\begin{figure}
\begin{centering}
\includegraphics[width=0.99\columnwidth]{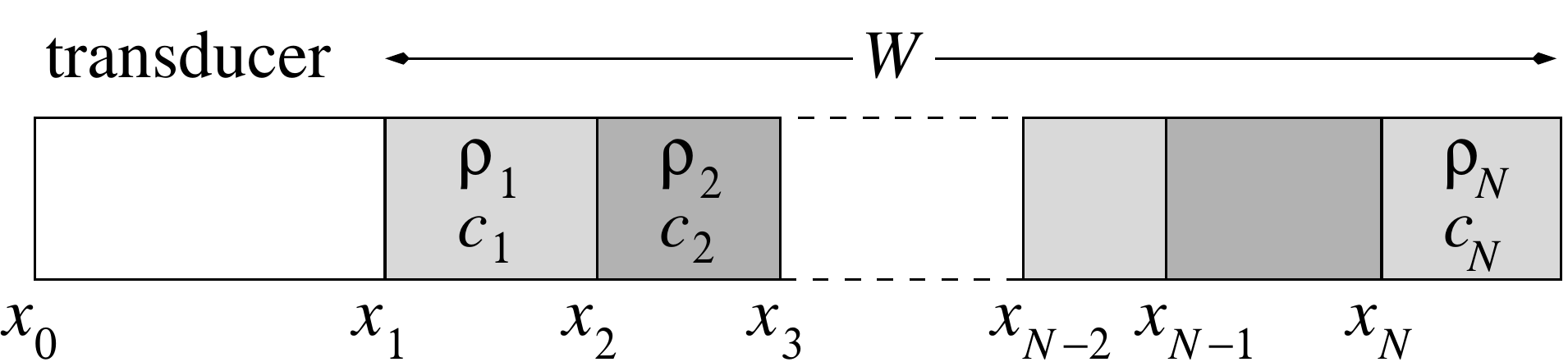}
\par\end{centering}

\caption{\label{fig:meiosec2}Sketch of the geometry used in simulating ultrasound
propagation in an inhomogeneous medium. See main text for labels.}
\end{figure}
The domains are labeled by an index $j$, so that domain $j$ extends
between $x_{j}$ and $x_{j+1}$, and is characterized by its density
$\rho_{j}$ and its sound velocity $c_{j}$. (See Figure \ref{fig:meiosec2}.)
For the one-dimensional geometry employed here, the solution of the
wave equation can be carried out semi-analytically, as detailed in
Appendix A. For a given choice of physical parameters $\left\{ \rho_{j},c_{j}\right\} $
of the various domains, the displacement field inside the medium,
as a function of position $x$ and time $t$, can be written as
\begin{equation}
\Phi\left(x,t\right)=\sum_{k}\phi_{k}X_{k}\left(x\right)\cos\left(\omega_{k}t\right),\label{eq:phixt}
\end{equation}
where $k$ labels the different eigenfrequencies $\omega_{k}$, the
function $X_{k}\left(x\right)$ is explicitly given by 
\begin{equation}
X_{k}\left(x\right)=A_{jk}\cos\left(\omega_{k}x/c_{j}\right)+B_{jk}\sin\left(\omega_{k}x/c_{j}\right),
\end{equation}
with $j$ such that $x_{j}\leq x<x_{j+1}$, and the coefficients $A_{jk}$
and $B_{jk}$ are determined from boundary conditions at the interfaces
separating contiguous domains, while the weights $\phi_{k}$ are derived
from the initial condition $\Phi\left(x,0\right)$. Here we choose
an initial pulse contained entirely inside the transducer, with a
form given by 
\begin{equation}
\Phi\left(x,0\right)=\Phi_{0}e^{-\gamma\left(x-x_{\mathrm{trans}}\right)}\sin\left[\frac{2\pi f\left(x-x_{\mathrm{trans}}\right)}{c_{\mathrm{trans}}}\right],
\end{equation}
for $x$ inside the transducer, and $\Phi\left(x,0\right)=0$ otherwise,
where $\Phi_{0}$ is an amplitude, $f$ is the reference frequency
of the transducer, $c_{\mathrm{trans}}$ is the sound velocity inside
the transducer, $x_{\mathrm{trans}}$ is the position of the left
end of the transducer, and $\gamma$ is a ``damping'' factor, introduced
so as to make the pulse resemble those produced by a real transducer.
In this work, we use $f=10\,\mathrm{MHz}$, $c_{\mathrm{trans}}=5800\,\mathrm{m/s}$,
and $\gamma=129.31\,\mathrm{m}^{-1}$, for a transducer of length
$2.32\,\mathrm{cm}$, in which 4 wavelengths of the pulse can fit.
The density inside the transducer is chosen as $\rho_{\mathrm{trans}}=2600\,\mathrm{kg/m}^{3}$
(about the density of quartz). We take into account in Eq. (\ref{eq:phixt})
all eigenfrequencies $\omega_{k}$ smaller than $\omega_{\mathrm{max}}=16\times2\pi f$,
corresponding to $16$ times the reference angular frequency of the
transducer. We checked that halving the value of $\omega_{\mathrm{max}}$
has no relevant effect on the results we report below.

From the displacement field, the sound pressure increments can be
calculated as
\begin{equation}
p\left(x,t\right)=\rho_{j}c_{j}^{2}\frac{\partial\Phi\left(x,t\right)}{\partial x}=\rho_{j}c_{j}^{2}\sum_{k}\phi_{k}X_{k}^{\prime}\left(x\right)\cos\left(\omega_{k}t\right),
\end{equation}
again with $j$ such that $x_{j}\leq x<x_{j+1}$. The ultrasound signals
we keep correspond to time series of the displacement and the pressure
increments captured at the right end of the transducer, with a sampling
rate of $50\,\mathrm{MHz}$. Each time series contains $2048$ points,
corresponding to about $4\times10^{-5}$ seconds. 

\begin{table}
\begin{centering}
\begin{tabular}{ccc}
\hline 
Microstructure & Average domain size ($\mathrm{m}$) & Average density ($\mathrm{kg}/\mathrm{m}^{3}$)\tabularnewline
\hline 
1 & $1.4\times10^{-5}$ & $7900$\tabularnewline
2 & $5.6\times10^{-5}$ & $7900$\tabularnewline
3 & $2.24\times10^{-4}$ & $7900$\tabularnewline
4 & $8.96\times10^{-4}$ & $7900$\tabularnewline
5 & $1.4\times10^{-5}$ & $6959$\tabularnewline
6 & $5.6\times10^{-5}$ & $6959$\tabularnewline
7 & $2.24\times10^{-4}$ & $6959$\tabularnewline
8 & $8.96\times10^{-4}$ & $6959$\tabularnewline
9 & $1.4\times10^{-5}$ & $6130$\tabularnewline
10 & $5.6\times10^{-5}$ & $6130$\tabularnewline
11 & $2.24\times10^{-4}$ & $6130$\tabularnewline
12 & $8.96\times10^{-4}$ & $6130$\tabularnewline
13 & $1.4\times10^{-5}$ & $5400$\tabularnewline
14 & $5.6\times10^{-5}$ & $5400$\tabularnewline
15 & $2.24\times10^{-4}$ & $5400$\tabularnewline
16 & $8.96\times10^{-4}$ & $5400$\tabularnewline
\hline 
\end{tabular}
\par\end{centering}

\caption{\label{tab:domaintypes}Average values of physical properties for
the 16 different microstructures used in this work. The sound velocity
is fixed at $5900\,\mathrm{\mathrm{m}/\mathrm{s}}$ in all cases.}
\end{table}
\begin{figure}
\begin{centering}
\includegraphics[width=0.99\columnwidth]{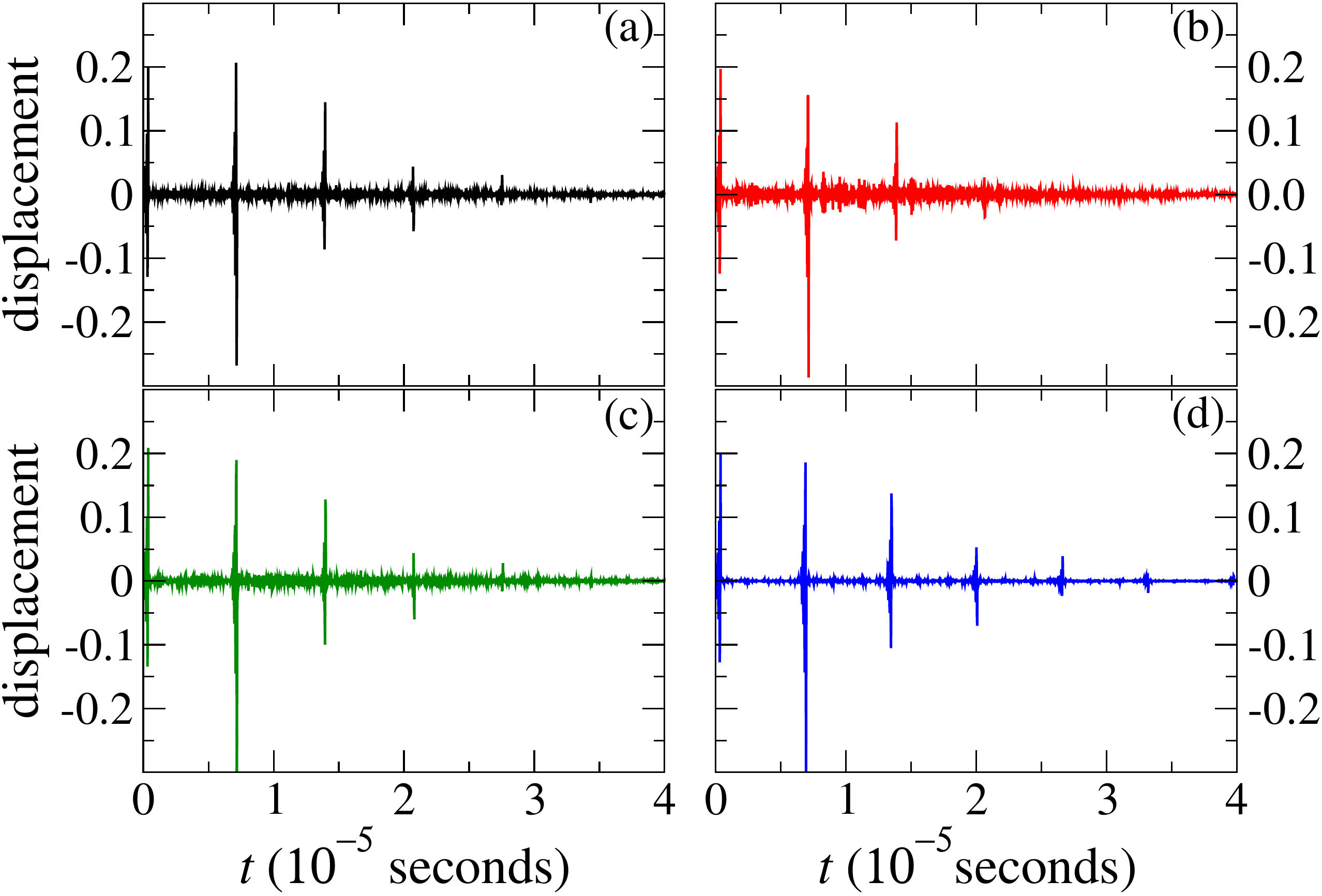}
\par\end{centering}

\caption{Typical simulated ultrasound signals from microstructures $1$ (a)
to $4$ (d).\label{fig:sampleascans}}
\end{figure}
We work here with 16 different choices of microstructure, combining
4 different average domain sizes and 4 different average densities,
with characteristics detailed in Table \ref{tab:domaintypes}. The
actual size and density of a domain are chosen from Gaussian probability
distributions with a standard deviation of 10\% the average value
for the size, and of 2.5\% the average value for the density. For
each of the $16$ microstructures, we obtained signals from $100$
different disorder realizations, with a total of $1600$ signals.
Our aim is to identify the microstructure based on the analysis of
the ultrasound signal. Since we keep the average system size fixed
at $W=2\,\mathrm{cm}$, we assume the same sound velocity ($5900\,\mathrm{\mathrm{m}/\mathrm{s}}$)
for all microstructures, so that the time intervals between signal
peaks do not trivially reveal the microstructure type. Typical signals
are shown in Fig. \ref{fig:sampleascans}. Notice that fluctuations
in the signals increase from microstructure $1$ to microstructure
$2$, and decrease for microstructures $3$ to $4$. This nonmonotonic
behavior of the fluctuations as a function of average domain size
is due to the fact that the average time needed for the wave to propagate
through a domain in microstructure $2$ is about $1/f=10^{-5}\,\mathrm{s}$,
thus maximizing scattering.

\section{Detrendend fluctuation analysis}

\label{sec:dfa}The detrended fluctuation analysis (DFA), introduced
by Peng \emph{et al.} \cite{peng94}, calculates the detrended fluctuations
in a time series as a function of a time-window size $\tau$. The
detrending comes from fitting the integrated time series inside each
time window to a polynomial, and calculating the average variance
of the residuals. Explicitly, the method works as follows. A time
series $\left\{ u_{i}\right\} $ of length $M$ is initially integrated,
yielding a new time series $y_{j}$, 
\begin{equation}
y_{j}=\sum_{i=1}^{j}\left(u_{i}-\left\langle u\right\rangle \right),\label{eq:dfaint}
\end{equation}
the average $\left\langle u\right\rangle $ being taken over all points,
\begin{equation}
\left\langle u\right\rangle =\frac{1}{M}\sum_{i=1}^{M}u_{i}.
\end{equation}
For each time window $I_{k}$ of size $\tau$, the points inside $I_{k}$
are fitted by a polynomial of degree $l$ (which we take in this work
to be $l=1$, i.e. a straight line), yielding a local trend $\tilde{y}_{j}$,
corresponding to the ordinate of the fit. The variance of the residuals
is calculated as 
\begin{equation}
f_{k}^{2}\left(\tau\right)=\frac{1}{\tau-1}\sum_{i\in I_{k}}\left(y_{j}-\tilde{y}_{j}\right)^{2},
\end{equation}
and $f_{k}\left(\tau\right)$ is averaged over all intervals to yield
the detrended fluctuation $F\left(\tau\right)$,
\begin{equation}
F\left(\tau\right)=\frac{1}{M-\tau+1}\sum_{k}f_{k}\left(\tau\right),
\end{equation}
$M-\tau+1$ being the number of time windows of size $\tau$ in a
time series with $M$ points. As defined here, $\tau$ is the (integer)
number of points inside a time window, the time increment between
consecutive points corresponding to the inverse sampling rate, $2\times10^{-8}\,\mathrm{s}$.

Notice that here, besides using overlapping time windows, we also
calculate the variance of the residuals inside each window, in a similar
spirit to what is done for the detrended cross-correlation analysis
of Ref. \cite{podobnik08}. This approach is slightly distinct from
the original scheme of Ref. \cite{peng94}, where nonoverlapping time
windows are employed, and the variance is calculated for the whole
time series. When applied to fractional Brownian motion \cite{addison97}
characterized by a Hurst exponent $H$, both approaches yield the
same exponent within numerical errors. Interestingly, the performance
of the classifier described in Sec. \ref{sec:gaussiandiscriminants},
however, is significantly improved by our approach.

When applied to a time series generated by a process governed by a
single dynamics, as for instance in fractional Brownian motion \cite{addison97},
DFA yields a function $F\left(\tau\right)$ following a power-law
behavior,
\begin{equation}
F\left(\tau\right)\sim C\tau^{\alpha},
\end{equation}
in which $C$ is a constant and $\alpha$ is an exponent which is
related to the Hurst exponent $H$, measuring memory effects on the
dynamics. If persistent (antipersistent) behavior of the time series
is favored, $\alpha$ is larger (smaller) than $1/2$.

\begin{figure}
\centering{}\includegraphics[width=0.99\columnwidth]{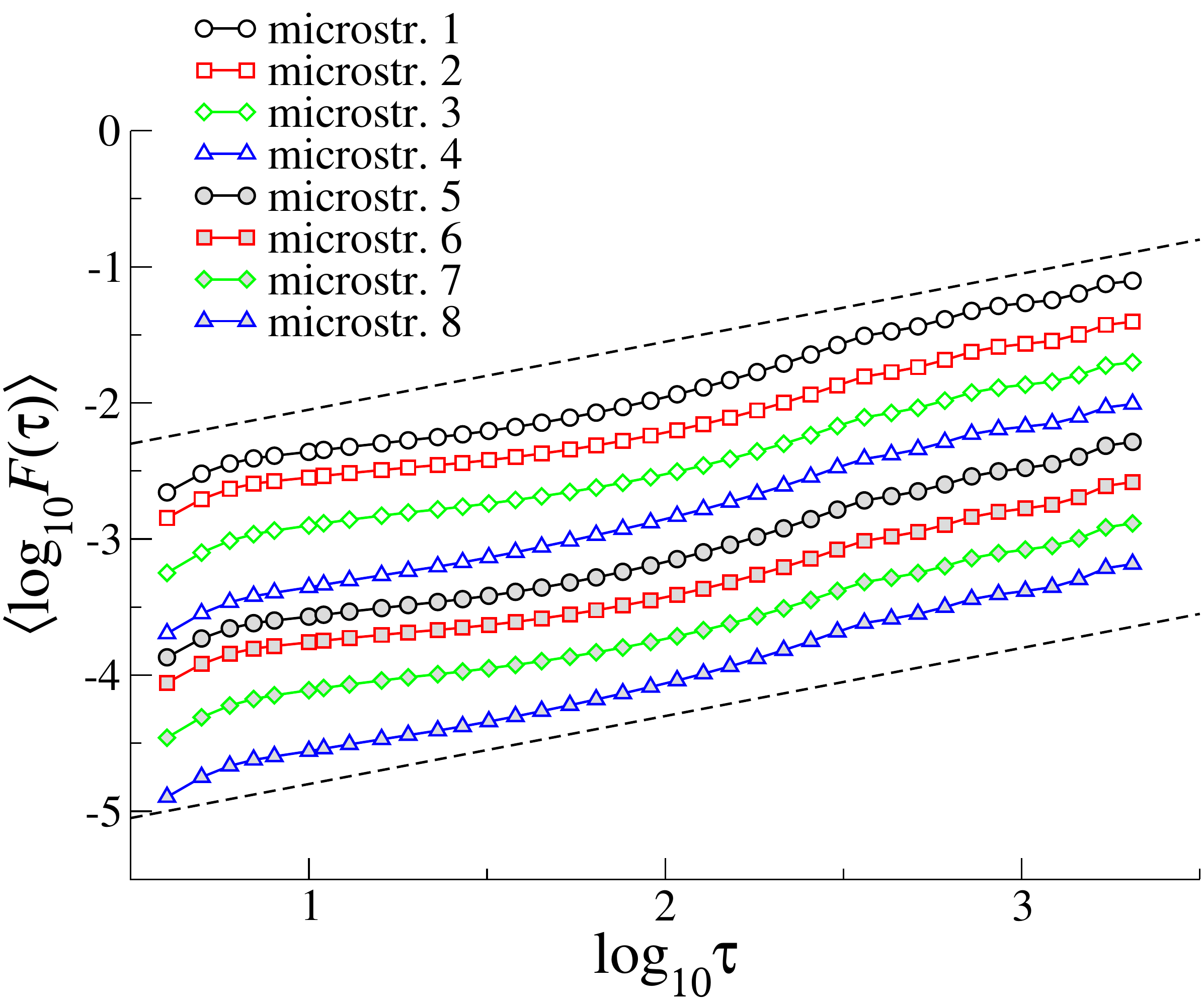}\caption{Average DFA curves for 8 different microstructures as a function of
the the time-window size $\tau$ (offset vertically for clarity).
The dashed lines have slope $1/2$.\label{fig:averagecurves}}
\end{figure}
As shown in Fig. \ref{fig:averagecurves}, for a subset of $8$ microstructures,
the curves of $F\left(\tau\right)$ calculated from the displacement
fields of simulated ultrasonic signals do not conform in general to
a power-law behavior, so that the exponent $\alpha$ is ill-defined,
except perhaps for microstructures $4$ and $8$, characterized by
the largest average domain sizes, which yield exponents approaching
the uncorrelated random-walk value $\alpha=1/2$. This same value
can be approximately identified for the other microstructures if the
analysis is restricted to a range of time-window sizes $\tau$ such
that $\log_{10}\tau\geq2.5$, which correspond to time scales greater
than $6.3\times10^{-6}\,\mathrm{s}$, compatible with the time $6.78\times10^{-6}\,\mathrm{s}$
needed for the pulse to travel across the medium and return to the
transducer. At shorter time scales, scattering of the waves at the
interfaces between domains introduces large interference effects leading
to the antipersistent behavior revealed by the $F\left(\tau\right)$
curves. Such effects, as expected, are stronger for microstructures
$1$, $2$, $5$, and $6$, characterized by smaller average domain
sizes. Even shorter time-window sizes, $\log_{10}\tau\lesssim0.7$,
probe time scales inferior to the inverse frequency of the pulse,
$1/f=10^{-7}\,\mathrm{s}$, and, as expected, point to a locally persistent
behavior of the time series.

Instead of attempting to correlate the signals with the microstructures
based on a manual identification of the various aspects of the curves,
we resort to pattern recognition tools \cite{webb02}. To this end,
we define for each signal $i$ a DFA vector $\mathbf{x}_{i}$ whose
components correspond to the values of the function $F\left(\tau\right)$
at a fixed set $\{\tau_{j}\}$ of window sizes. Here we build $\{\tau_{j}\}$
from the integer part of the integer powers of $2^{1/4}$, from $4$
to $2048$, comprising $37$ different values of $\tau$. 

\begin{figure}
\centering{}\includegraphics[width=0.99\columnwidth]{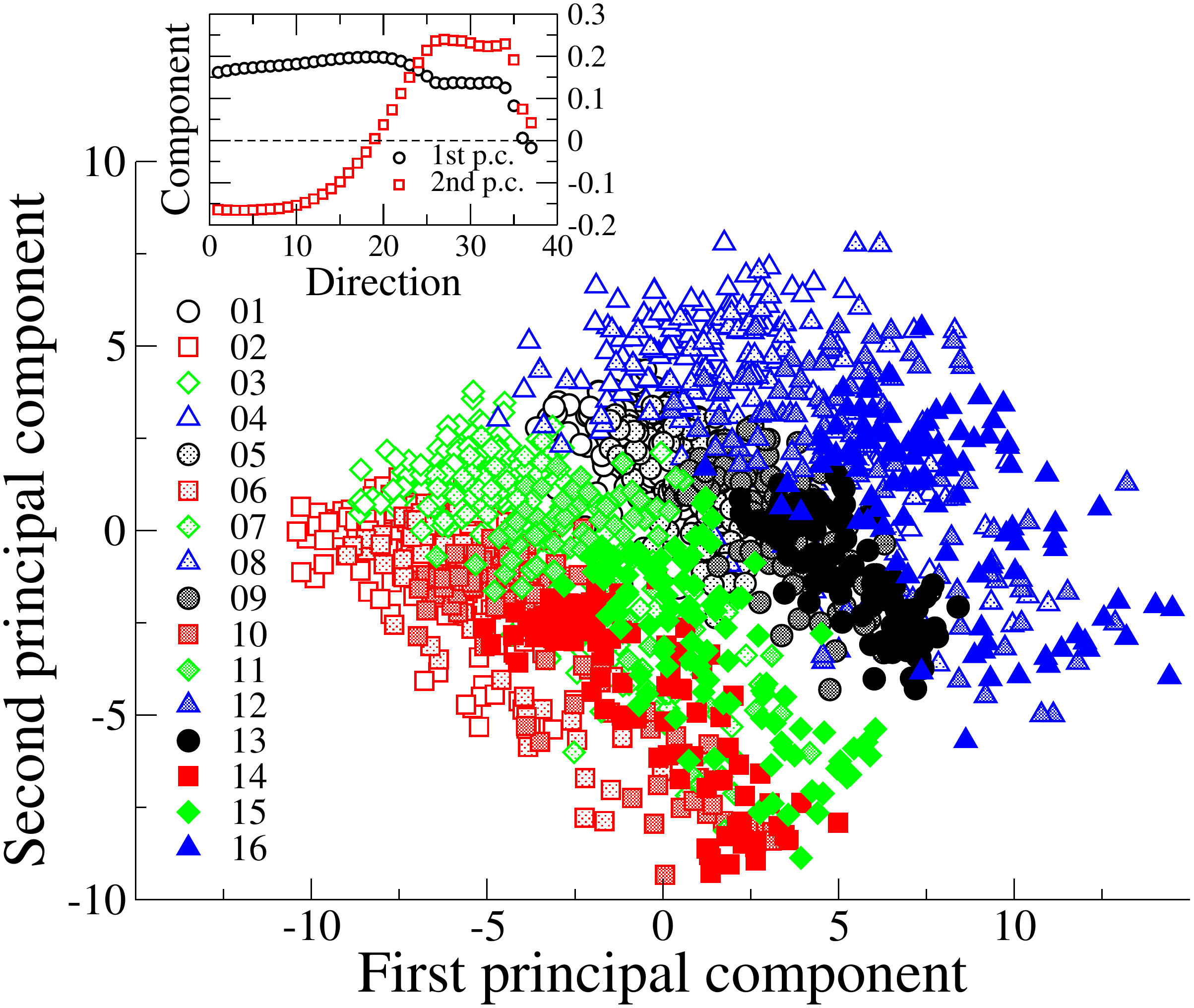}\caption{Projection of the DFA vectors on the plane defined by the first two
principal components of the data, corresponding to the eigenvectors
of the covariance matrix associated with the two largest eigenvalues.
Inset: vector components of the eigenvectors for each of the $37$
directions.\label{fig:pc1vspc2}}
\end{figure}
A visualization of the DFA vectors is hindered by their high number
of components, $n=37$. However, a principal-component analysis \cite{webb02}
can reveal the directions along which the data for all $1600$ vectors
is most scattered. This is done by projecting each vector along the
principal components, corresponding to the eigenvectors of the covariance
matrix
\begin{equation}
\boldsymbol{\Sigma}=\frac{1}{N}\sum_{i}\left(\mathbf{x}_{i}-\boldsymbol{\mu}\right)\left(\mathbf{x}_{i}-\boldsymbol{\mu}\right)^{T},
\end{equation}
in which the summation runs over all $N=1600$ DFA (column) vectors
$\mathbf{x}_{i}$, while $\boldsymbol{\mu}$ is the average vector,
\begin{equation}
\boldsymbol{\mu}=\frac{1}{N}\sum_{i}\mathbf{x}_{i}.
\end{equation}
The eigenvectors are arranged in decreasing order of their respective
eigenvalues. The first principal component thus corresponds to the
direction accounting for the largest variation in the data, the second
principal component to the second largest variation, etc. Figure \ref{fig:pc1vspc2}
shows projections of all DFA vectors for displacements on the plane
defined by the first two principal components, revealing at the same
time a clustering of the data for each microstructure and a considerable
superposition of the data for microstructures which differ only by
average density (microstructures $1$, $5$, $9$, and $13$, for
instance). Although this superposition is in part an artifact of the
two-dimensional projection, it is not satisfactorily eliminated when
the other directions are taken into account. Accordingly, attempts
at associating a vector $\mathbf{x}_{i}$ to the microstructure whose
average vector is closer to $\mathbf{x}_{i}$ lead to an error rate
of about $42\%$. However, as discussed in the next Section, a more
sophisticated approach considerably improves the classification performance. 

Notice that, as shown in the inset of Fig. \ref{fig:pc1vspc2}, special
features of the first two eigenvectors of the covariance matrix are
associated with directions $26$ and above, along which the components
of the first (second) eigenvector have typically smaller (larger)
absolute values than along other directions. Moreover, the components
of the second eigenvector along directions $1$ to $5$ also have
locally larger absolute values. In fact, direction $5$ is connected
with the inverse frequency $1/f$, and directions above $26$ are
associated with time-window sizes such that $\log_{10}\tau>2.5$,
again pointing to the special role played by these time scales in
differentiating the various microstructures.

\section{Gaussian discriminants}

\label{sec:gaussiandiscriminants}We first want to check whether it
is possible to build an efficient automated classifier which is able
to assign a signal to one of the microstructures, based on the corresponding
DFA vector. Attempts at assigning a DFA vector $\mathbf{x}$ to the
microstructure (or \emph{class}, in the language of pattern recognition)
whose average vector lies closer to $\mathbf{x}$ lead to many classifications
due to the fact that the average vectors of similar microstructures
(such as $1$ and $5$) are close to one another. As a classification
based solely on distance disregards additional information provided
by the probability distributions of the DFA vectors obtained from
each microstructure, whose variances along different directions can
exhibit different profiles, here we follow an approach to discrimination
based on estimates of those distributions.

Our task is to estimate the probability $P\left(\omega_{j}|\mathbf{x}\right)$
that a given vector $\mathbf{x}$ belongs to class $\omega_{j}$,
$j\in\left\{ 1,2,\ldots,C\right\} $. (In our case, of course, $C=16$.)
From Bayes' theorem, this probability can be written as
\begin{equation}
P\left(\omega_{j}|\mathbf{x}\right)=\frac{P\left(\mathbf{x}|\omega_{j}\right)P\left(\omega_{j}\right)}{P\left(\mathbf{x}\right)},
\end{equation}
where $P\left(\mathbf{x}|\omega_{j}\right)$ is the probability that
a sample belonging to class $\omega_{j}$ produces a vector $\mathbf{x}$,
$P\left(\omega_{j}\right)$ is the prior probability of class $\omega_{j}$
occurring, and $P\left(\mathbf{x}\right)$ is the prior probability
of vector $\mathbf{x}$ occurring. Once $P\left(\omega_{j}|\mathbf{x}\right)$
is known for all classes $\omega_{j}$, we assign vector $\mathbf{x}$
to class $\omega_{j}$ if
\[
P\left(\omega_{j}|\mathbf{x}\right)>P\left(\omega_{k}|\mathbf{x}\right),\quad\mbox{for all }k\neq j.
\]
Since $P\left(\mathbf{x}\right)$ is class-independent, and thus irrelevant
to the decision process, the problem of calculating $P\left(\omega_{j}|\mathbf{x}\right)$
reduces to estimating $P\left(\mathbf{x}|\omega_{j}\right)$ and $P\left(\omega_{j}\right)$.

Among the various possibilities for the estimation of $P\left(\mathbf{x}|\omega_{j}\right)$,
we choose to work with normal-based quadratic discriminant functions
\cite{webb02}, derived as follows. We assume that $P\left(\mathbf{x}|\omega_{j}\right)$
has the Gaussian form
\begin{equation}
P\left(\mathbf{x}|\omega_{j}\right)=\frac{1}{\left(2\pi\right)^{\frac{n}{2}}\left|\boldsymbol{\Sigma}_{j}\right|^{\frac{1}{2}}}\exp\left[-\frac{1}{2}\left(\mathbf{x}-\boldsymbol{\mu}_{j}\right)^{T}\boldsymbol{\Sigma}_{j}^{-1}\left(\mathbf{x}-\boldsymbol{\mu}_{j}\right)\right],
\end{equation}
where $n$ is the number of components of $\mathbf{x}$, while $\bm{\mu}_{j}$
and $\boldsymbol{\Sigma}_{j}$ are the average vector and the covariance
matrix of class $\omega_{j}$. By selecting a subset of the available
vectors to form a training set $\left\{ \mathbf{x}_{i}\right\} $,
unbiased maximum-likelihood estimates of $\boldsymbol{\mu}_{j}$ and
$\boldsymbol{\Sigma}_{j}$ are provided by
\begin{equation}
\hat{\boldsymbol{\mu}}_{j}=\frac{1}{\mathcal{N}_{j}}\sum_{i\in\omega_{j}}\mathbf{x}_{i}
\end{equation}
and
\begin{equation}
\boldsymbol{\hat{\Sigma}}_{j}=\frac{1}{\mathcal{N}_{j}-1}\sum_{i\in\omega_{j}}\left(\mathbf{x}_{i}-\hat{\boldsymbol{\mu}}_{j}\right)\left(\mathbf{x}_{i}-\hat{\boldsymbol{\mu}}_{j}\right)^{T},
\end{equation}
with $\mathcal{N}_{j}$ the number of vectors in the training set
belonging to class $\omega_{j}$. The decision process then corresponds
to assigning a vector $\mathbf{x}$ to class $\omega_{j}$ if $g_{j}\left(\mathbf{x}\right)>g_{k}\left(\mathbf{x}\right)$
for all $k\neq j$, where
\begin{equation}
g_{j}\left(\mathbf{x}\right)=\ln\hat{P}\left(\omega_{j}\right)-\frac{1}{2}\ln\left|\boldsymbol{\hat{\Sigma}}_{j}\right|-\frac{1}{2}\left(\mathbf{x}-\hat{\boldsymbol{\mu}}_{j}\right)^{T}\boldsymbol{\hat{\Sigma}}_{j}^{-1}\left(\mathbf{x}-\hat{\boldsymbol{\mu}}_{j}\right),
\end{equation}
an estimate of $P\left(\omega_{j}\right)$ being given by
\begin{equation}
\hat{P}\left(\omega_{j}\right)=\frac{\mathcal{N}_{j}}{\sum_{k=1}^{C}\mathcal{N}_{k}}.
\end{equation}

\begin{table}
\begin{tabular}{c>{\centering}p{0.2\columnwidth}>{\centering}p{0.45\columnwidth}}
\hline 
Microstructure & Success rate  & Misclassifications\tabularnewline
\hline 
1  & \textbf{98.6} (0.3)  & 5: 1.4\tabularnewline
2  & \textbf{98.7} (0.2)  & 6: 1.3\tabularnewline
3  & \textbf{99.6} (0.1) & 4: 0.1$\quad$ 7: 0.3\tabularnewline
4  & \textbf{99.0} (0.2)  & 8: 1.0\tabularnewline
5 & \textbf{99.1} (0.2) & 1: 0.4$\quad$9: 0.5\tabularnewline
6 & \textbf{99.8} (0.1) & 2: 0.1$\quad$10: 0.1\tabularnewline
7 & \textbf{98.8} (0.2) & 3: 1.13$\quad$ 8: 0.04$\quad$11: 0.03\tabularnewline
8 & \textbf{97.0} (0.4) & 4: 2.9$\quad$12: 0.1\tabularnewline
9 & \textbf{99.5} (0.1) & 5: 0.5\tabularnewline
10 & \textbf{99.8} (0.1) & 14: 0.2\tabularnewline
11 & \textbf{99.6} (0.1) & 7: 0.3$\quad$15: 0.1\tabularnewline
12 & \textbf{99.4} (0.2) & 8: 0.6\tabularnewline
13 & \textbf{99.7} (0.1) & 9: 0.3\tabularnewline
14 & \textbf{100} & none\tabularnewline
15 & \textbf{100} & none\tabularnewline
16 & \textbf{99.1} (0.2) & 12: 0.9\tabularnewline
\hline 
\end{tabular}\caption{Average performance of the classifier when applied to the testing
vectors built from displacements, calculated over $100$ sets of training
and testing vectors. For the second column, bold numbers indicate
the percentage of vectors which were correctly classified, figures
in parenthesis corresponding to the standard deviations. The third
column registers average misclassification rates (in the first row,
`5: 1.4' indicates that 1.4\% of vectors belonging to microstructure
1 were misclassified as belonging to microstructure 5).\label{tab:confmatrix}}
\end{table}
First we tested the classifier by using all the $1600$ DFA vectors
for displacements as the training set. This yields functions $g_{k}\left(\mathbf{x}\right)$
that are able to correctly classify all vectors, a flawless performance.
In order to evaluate the generalizability of the classifier results,
we randomly selected $80\%$ ($1280$) of the $1600$ available vectors
to define the training set, using the remaining vectors in the testing
stage, and took averages over $100$ distinct choices of training
and testing sets. When the training vectors were fed back to the classifier,
as a first step toward validation, again no vectors were misclassified.
Table \ref{tab:confmatrix} summarizes the average performance of
the classifier when applied to the testing vectors built from the
DFA analysis of the displacement signals. Notice that the maximum
average classification error corresponds to $3\%$, for microstructure
$8$. The overall testing error, taking into account all classes,
corresponds to $0.8\%$. Misclassifications almost exclusively involve
assigning a vector to a microstructure with the correct average domain
size but different average density (e.g. microstructures $4$ or $12$
instead of $8$), with very few cases involving the same density but
with a different although closest average domain size (microstructure
$8$ instead of $7$). This fact can be used to build a classifier
which groups vectors into 4 classes, according to the average domain
size of the corresponding microstructures, with no misclassifications.
A similar classifier targeting average densities rather than average
domain sizes shows only a very small misclassification rate of $0.01\%$.

Interestingly, processing the displacement signals according to the
original DFA recipe of Ref. \cite{peng94} leads to an inferior performance
for microstructure classification, with an average error of $29\%$,
but now most errors involve vectors being assigned to classes with
the same density as that of the correct microstructure. A classifier
which groups vectors according to the average density of the corresponding
microstructures achieves a misclassification rate of only $2\%$.

The efficiency of the classifiers is dependent on the choice of values
of the time-window sizes. For instance, in the 16-class case, restricting
the values of $\tau_{j}$ to the powers of $2$ doubles the overall
testing error, to around $1.7\%$, while expanding $\left\{ \tau_{j}\right\} $
to the integer parts of the powers of $2^{1/8}$ leads to a much larger
overall testing error of $24\%$. Choosing $\left\{ \tau_{j}\right\} $
as the integer parts of the powers of $2^{1/2}$ actually leads to
a slightly smaller overall testing error of $0.6\%$, but a few training
errors also occur. Thus, our choice of $\left\{ \tau_{j}\right\} $
from the integer parts of the powers of $2^{1/4}$ seems to be close
to optimal. In contrast, performing the detrended fluctuation analysis
according to the original recipe of Ref. \cite{peng94} leads to a
minimum overall testing error of $28\%$ as the values of the time-window
sizes are varied. 

\begin{table}
\begin{tabular}{c>{\centering}p{0.2\columnwidth}>{\centering}p{0.45\columnwidth}}
\hline 
Microstructure & Success rate  & Misclassifications\tabularnewline
\hline 
1  & \textbf{97.6} (0.4)  & 5: 1.6\tabularnewline
2  & \textbf{83.5} (1.0)  & 6: 16.2\tabularnewline
3  & \textbf{88.9} (0.8) & 7: 11.0\tabularnewline
4  & \textbf{91.7} (0.6)  & 8: 8.2\tabularnewline
5 & \textbf{93.6} (0.6) & 1: 1.2$\quad$6:2.7$\quad$9: 2.3\tabularnewline
6 & \textbf{80.9} (1.1) & 2: 10.5$\quad$10: 8.5\tabularnewline
7 & \textbf{81.6} (1.0) & 3: 12.2$\quad$ 11: 5.2\tabularnewline
8 & \textbf{92.0} (0.7) & 4: 5.1$\quad$12: 2.8\tabularnewline
9 & \textbf{93.9} (0.6) & 5: 3.3$\quad$10: 2.5\tabularnewline
10 & \textbf{85.7} (0.9) & 6: 9.3$\quad$14: 4.6\tabularnewline
11 & \textbf{86.5} (0.9) & 7: 8.6$\quad$15: 4.8\tabularnewline
12 & \textbf{86.3} (0.9) & 8: 11.2$\quad$ 16: 2.3\tabularnewline
13 & \textbf{99.9} (0.1) & 9: 0.1\tabularnewline
14 & \textbf{95.2} (0.5) & 10: 4.8\tabularnewline
15 & \textbf{94.3} (0.6) & 11: 5.5\tabularnewline
16 & \textbf{92.7} (0.6) & 12: 7.0\tabularnewline
\hline 
\end{tabular}\caption{The same as in Table \ref{tab:confmatrix}, but now with testing vectors
built from pressure increments. Except for class 13, only misclassification
rates above 1\% are shown. \label{tab:confmatrix-pressure}}
\end{table}
Processing the pressure signals according to the DFA scheme of Sec.
\ref{sec:dfa} leads to a decrease in the performance of the classifier,
which is partially recovered by omitting the initial integration of
the signal, prescribed by Eq. \ref{eq:dfaint}. The results are shown
in Table \ref{tab:confmatrix-pressure}. The overall classification
error now corresponds to $10\%$, and once more most errors involve
assigning vectors to classes with the correct average domain size
but different densities. A classifier grouping the vectors according
to the average domain size shows a misclassification rate of only
$0.14\%$, while the analogous classifier based on average densities
yields an error rate of $4\%$. Again, using the original DFA scheme
(with no initial integration of the signal) increases the overall
error rate for 16 classes (to $23\%$), but interchanges the performances
of classifiers aiming only at domain sizes (error rate of $13\%$)
or densities (error rate of $5\%$).

\section{Neural networks}

\label{sec:neuralnet}In the spirit of Refs. \cite{ovchinnikov2009}
and \cite{kumar2011}, which employed artificial neural networks in
order to identify disorder parameters in the random-bond random-field
Ising model, we wish to investigate whether a similar approach can
be useful in estimating average domain sizes and average densities
based on fluctuation analyses of our simulated ultrasound data.

The idea here is to build a neural network which reads the DFA vectors
as inputs, targets as outputs the physical parameters (either average
domain size or average density) from all vectors of 15 of the 16 possible
microstructures, and then tries to guess the corresponding parameter
from the DFA vectors of the remaining class. The network --- a multilayer
perceptron \cite{rumelhart1986,duda2000} --- is composed of an input
layer with $N_{1}=37$ neurons, which receive the data from each DFA
vector, an output layer with a single neuron ($N_{4}=1$), whose reading
is the desired physical parameter, and two hidden layers, containing
respectively $N_{2}=18$ and $N_{3}=8$ neurons. The connection weights
between neurons in contiguous layers are adjusted so as to minimize
the mean square error between the desired and the actual outputs,
according to the backpropagation prescription. We employed the hyperbolic
tangent as the activation function %
\footnote{Precisely, the activation function employed was $a\tanh\left(bx\right)$,
with the recommended values $a=1.7159$ and $b=2/3$ \cite{haykin99}.%
}, and both input and output data were converted to a logarithmic scale
and adjusted so as to lie between $-1+\epsilon$ and $1-\epsilon$,
with $\epsilon$ of order $1$/10. (This rescaling improves the performance
of the perceptron when dealing with microstructures for which parameters
take extreme values.) In all cases, the networks were trained for
$3\,000$ epochs. 

\begin{figure}
\centering{}\includegraphics[width=0.99\columnwidth]{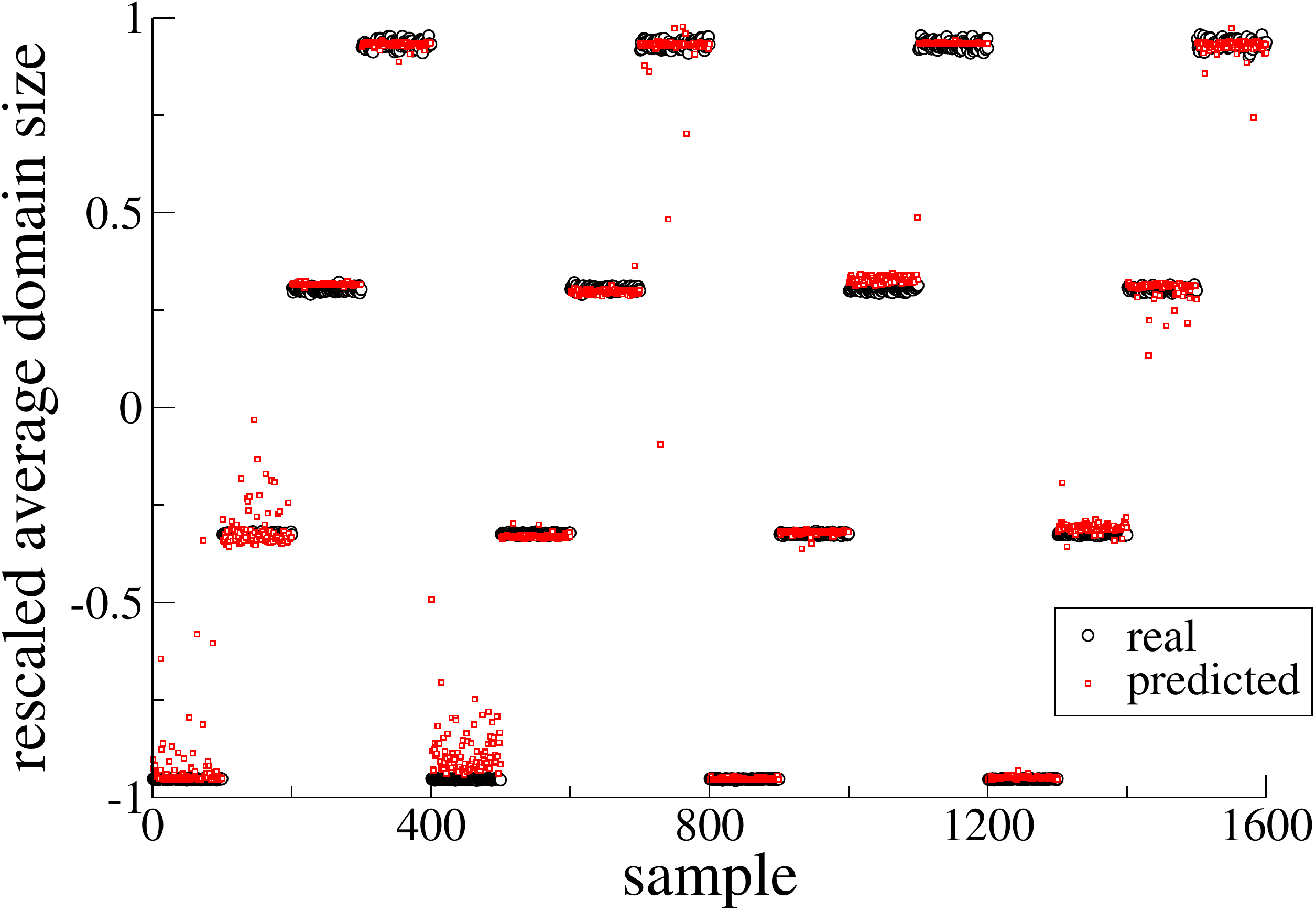}\caption{Real (circles) and predicted (squares) values of the rescaled average
domain sizes for different samples (numbered $1$ to $1600$).\label{fig:NNclasseexcluida-tam}}
\end{figure}
\begin{figure}
\centering{}\includegraphics[width=0.99\columnwidth]{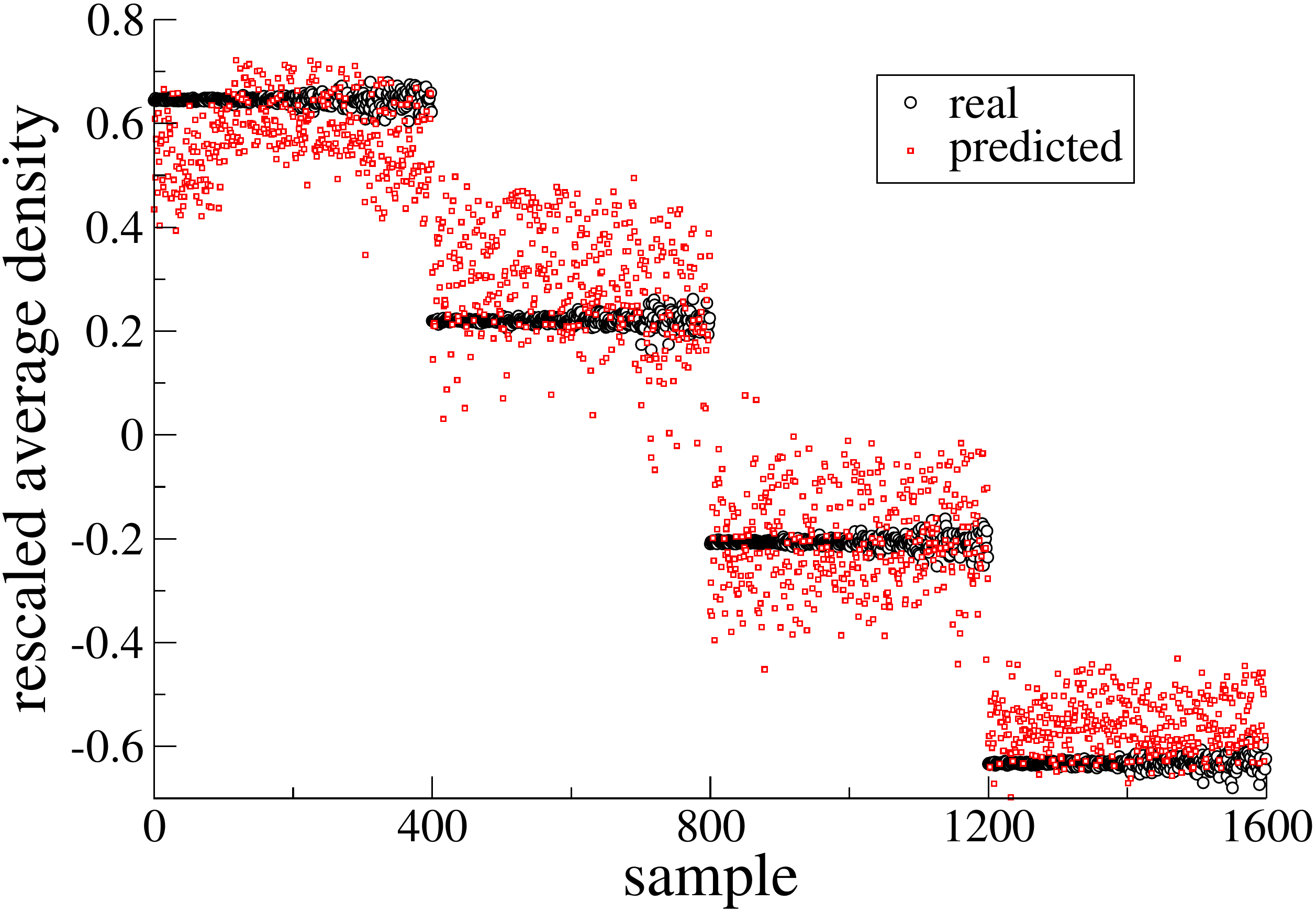}\caption{Real (circles) and predicted (squares) values of the rescaled average
densities for different samples (numbered $1$ to $1600$).\label{fig:NNclasseexcluida-dens}}
\end{figure}
Figures \ref{fig:NNclasseexcluida-tam} and \ref{fig:NNclasseexcluida-dens}
show plots of the rescaled average domain size and rescaled average
densities for each microstructure, along with the predictions output
when the perceptron is trained with displacement DFA data from all
remaining microstructures. Despite the relatively high average error
(of $0.01$ for the domain sizes and $0.1$ for the densities), it
is clear that the information hidden in the DFA vectors is enough
to provide useful predictions for the unknown parameters. (Using data
from pressure DFA vectors leads to a similar performance when predicting
densities, but a five-fold increase in the average error when predicting
domain sizes.) We also trained the perceptron to output both domain
size and density, but the performance showed a considerable decline
compared to the case when the parameters where targeted by different
networks. An early-stopping criterion (see e.g. Ref. \cite{haykin99})
was likewise implemented, but did not lead to improved results.

Finally, if the network is trained with a random selection of 1280
(80\%) samples from \emph{all} classes, the overall error in the testing
stage is reduced to around $10^{-3}$ when targeting the average logarithmic
domain size or average logarithmic density, indicating that this setup
can also be used as a classifier, in the same spirit as the Gaussian
discriminants of Sec. \ref{sec:gaussiandiscriminants}, although at
a considerably higher computational cost.

\section{Summary}

\label{sec:conclusions}Our aim in this work was to provide, within
a controllable framework, a proof-of-principle for the identification
of microstructures based on fluctuation analyses of ultrasound signals.
With a slightly modified detrended-fluctuation analysis (DFA) algorithm,
we were able to build an automated Gaussian classifier capable of
assigning a DFA curve to the correct microstructure among 16 possibilities,
corresponding to combinations of 4 average densities and 4 average
domain sizes, with an error rate below 1\%. Although not detailed
here, an analogous classifier based on the original DFA algorithm
of Ref. \cite{peng94}, despite not providing a comparable performance
(yielding a much larger error rate around 30\%), is able to separate
the microstructures according to their average densities with an error
rate of about 2\%. Incidentally, yet another analogous classifier
based on Hurst's $R/S$ analysis also performs modestly for overall
classification, with an error rate of about 22\%, but is able to separate
the microstructures according to their average domain size with a
success rate in excess of 99.7\%.

We also described a multilayer perceptron which is able to provide
estimates of a physical property for DFA curves from an unknown microstructure
after being trained to output the corresponding property for the remaining
microstructures.

The application of the methods described here to more realistic situations
depends on a series of tests which incorporate effects coming from
more complicated, higher-dimensional geometries. Among these effects,
we mention mode conversion at domain interfaces and the presence of
additional defects such as voids or inclusions of distinct phases.
We hope the results reported in this paper will encourage future investigations.
\begin{acknowledgments}
This work has been financed by the Brazilian agencies CNPq, FUNCAP,
and FINEP (CTPetro). A. P. Vieira acknowledges financial support from
Universidade de São Paulo (NAP-FCx) and FAPESB/PRONEX.
\end{acknowledgments}
\appendix

\section{Solution of the wave equation for a one-dimensional heterogeneous
medium}

\label{sec:appendix}
\begin{figure}
\begin{centering}
\includegraphics[width=0.99\columnwidth]{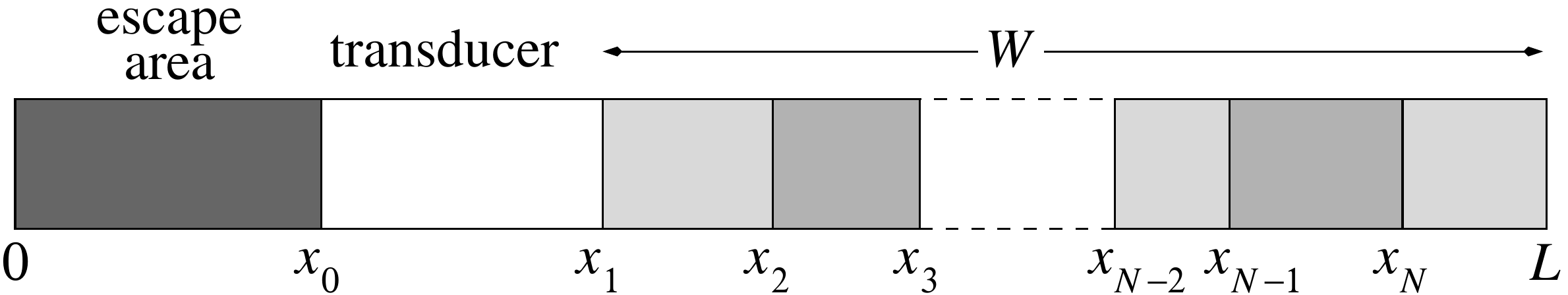}
\par\end{centering}

\caption{\label{fig:meioapp}Sketch of the geometry used in simulating ultrasound
propagation in an inhomogeneous medium. See main text for labels.}
\end{figure}
We want to solve the wave equation for the displacement field $\Phi\left(x,t\right)$,
\begin{equation}
\frac{\partial^{2}\Phi}{\partial x^{2}}=\frac{1}{c^{2}}\frac{\partial^{2}\Phi}{\partial t^{2}},
\end{equation}
with the sound velocity $c$ a constant along each domain into which
the one-dimensional system, of size $L$, is divided. This means that
$\Phi\left(x,t\right)$ will be given by a different function in each
domain, and the problem can be recast as the solution of the wave
equations
\begin{equation}
\frac{\partial^{2}\Phi_{j}}{\partial x^{2}}=\frac{1}{c_{j}^{2}}\frac{\partial^{2}\Phi_{j}}{\partial t^{2}},\label{eq:waveequationj}
\end{equation}
for every domain $j$, subject to the boundary conditions
\begin{equation}
\Phi_{j-1}\left(x_{j},t\right)=\Phi_{j}\left(x_{j},t\right)\label{eq:bcphi}
\end{equation}
and
\begin{equation}
\rho_{j-1}c_{j-1}^{2}\frac{\partial\Phi_{j-1}}{\partial x}\left(x_{j},t\right)=\rho_{j}c_{j}^{2}\frac{\partial\Phi_{j}}{\partial x}\left(x_{j},t\right),\label{eq:bcdphi}
\end{equation}
describing the continuity of the displacement and the pressure fields
at the interfaces between domains. Here, $c_{j}$ and $\rho_{j}$
denote the sound velocity and the density in domain $j$, while $x_{j}$
is the coordinate of the left end of domain $j$. The medium is divided
into $N+2$ domains $\left(j\in\left\{ -1,0,1,2,\ldots,N\right\} \right)$,
with $x_{-1}=0$ and $x_{N+1}=L$; see Figure \ref{fig:meioapp}.
Domains from $j=1$ to $j=N$ correspond to the medium to be investigated,
and span a length $W<L$. Domain $j=0$ holds a piezoelectric transducer,
in which the ultrasonic pulse is to be produced, and domain $j=-1$
is reserved for an ``escape area'', introduced to mimic the presence
of an absorbing wall at the back of the transducer. 

Separation of variables leads to a general solution of Eq. (\ref{eq:waveequationj}),
for a given angular frequency $\omega$, of the form
\begin{eqnarray}
\Phi_{j}\left(x,t;\omega\right) & = & \left[A_{j}\cos\left(k_{j}x\right)+B_{j}\sin\left(k_{j}x\right)\right]\cos\left(\omega t\right)\nonumber \\
 & + & \left[C_{j}\cos\left(k_{j}x\right)+D_{j}\sin\left(k_{j}x\right)\right]\sin\left(\omega t\right).
\end{eqnarray}
Since we will impose initial conditions in which $\frac{\partial\Phi_{j}}{\partial t}\equiv0$,
we can set $C_{j}=D_{j}=0$ for all $j$. The boundary conditions
in Eq. (\ref{eq:bcphi}) and (\ref{eq:bcdphi}) lead to
\begin{equation}
\frac{A_{j-1}\cos\left(k_{j-1}x_{j}\right)+B{}_{j-1}\sin\left(k_{j-1}x_{j}\right)}{A_{j}\cos\left(k_{j}x_{j}\right)+B{}_{j}\sin\left(k_{j}x_{j}\right)}=1\label{eq:interfacebc}
\end{equation}
and
\begin{equation}
\frac{z_{j-1}\left[A_{j-1}\sin\left(k_{j-1}x_{j}\right)-B{}_{j-1}\cos\left(k_{j-1}x_{j}\right)\right]}{z_{j}\left[A_{j}\sin\left(k_{j}x_{j}\right)-B{}_{j}\cos\left(k_{j}x_{j}\right)\right]}=1\label{eq:interfacebcz}
\end{equation}
for $j\in\left\{ 0,1,2,\ldots,N\right\} ,$ in which we introduced
the acoustic impedances $z_{j}=\rho_{j}c_{j}$, while the reflective
boundary conditions at $x=0$ and $x=L$, $\Phi_{-1}\left(0,t\right)=\Phi_{N}\left(L,t\right)=0$,
yield
\begin{equation}
\left\{ \begin{array}{l}
A_{-1}=0,\\
A_{N}\cos\left(k_{N}L\right)+B_{N}\sin\left(k_{N}L\right)=0.
\end{array}\right.\label{eq:reflectbc}
\end{equation}
In order to mimic an absorbing wall at the back of the transducer
(at $x=x_{0}$), we choose for domain $j=-1$ an acoustic impedance
$z_{-1}=z_{0}$ and a very small sound velocity $c_{-1}$. These choices
guarantee that waves incident on the left end of the transducer are
not reflected, rather entering the escape area and not returning during
the simulation.

Equations (\ref{eq:interfacebc}), (\ref{eq:interfacebcz}) and (\ref{eq:reflectbc})
constitute a homogeneous system of linear equations in the coefficients
$A_{j}$ and $B_{j}$. Rewriting the system as the matrix equation
\begin{equation}
\left(\begin{array}{ccccccc}
A_{-1} & B_{-1} & A_{0} & B_{0} & \cdots & A_{N} & B_{N}\end{array}\right)\mathbf{M}_{N}=0,
\end{equation}
in which\begin{widetext} 
\begin{equation}
\mathbf{M}_{N}=\left[\begin{array}{ccccccccc}
1 & \cos\left(k_{-1}x_{1}\right) & z_{-1}\sin\left(k_{-1}x_{1}\right) & 0 & 0 & \cdots & 0 & 0 & 0\\
0 & \sin\left(k_{-1}x_{1}\right) & -z_{-1}\cos\left(k_{-1}x_{1}\right) & 0 & 0 & \cdots & 0 & 0 & 0\\
0 & -\cos\left(k_{0}x_{1}\right) & -z_{0}\sin\left(k_{0}x_{1}\right) & \cos\left(k_{0}x_{1}\right) & z_{0}\sin\left(k_{0}x_{1}\right) & \cdots & 0 & 0 & 0\\
0 & -\sin\left(k_{0}x_{1}\right) & z_{0}\cos\left(k_{0}x_{1}\right) & \sin\left(k_{0}x_{1}\right) & -z_{0}\cos\left(k_{0}x_{1}\right) & \cdots & 0 & 0 & 0\\
0 & 0 & 0 & -\cos\left(k_{1}x_{1}\right) & -z_{1}\sin\left(k_{1}x_{1}\right) & \cdots & 0 & 0 & 0\\
0 & 0 & 0 & -\sin\left(k_{1}x_{1}\right) & z_{1}\cos\left(k_{1}x_{1}\right) & \cdots & 0 & 0 & 0\\
\vdots & \vdots & \vdots & \vdots & \vdots & \ddots & \vdots & \vdots & \vdots\\
0 & 0 & 0 & 0 & 0 & \cdots & \cos\left(k_{N-1}x_{N}\right) & z_{N-1}\sin\left(k_{N-1}x_{N}\right) & 0\\
0 & 0 & 0 & 0 & 0 & \cdots & \sin\left(k_{N-1}x_{N}\right) & -z_{N-1}\cos\left(k_{N-1}x_{N}\right) & 0\\
0 & 0 & 0 & 0 & 0 & \cdots & -\cos\left(k_{N}x_{N}\right) & -z_{N}\sin\left(k_{N}x_{N}\right) & \cos\left(k_{N}L\right)\\
0 & 0 & 0 & 0 & 0 & \cdots & -\sin\left(k_{N}x_{N}\right) & z_{N}\cos\left(k_{N}x_{N}\right) & \sin\left(k_{N}L\right)
\end{array}\right],
\end{equation}
\end{widetext}we see that nontrivial solutions for the $\left\{ A_{j},B_{j}\right\} $
are obtained only if
\[
\det\mathbf{M}_{N}=0,
\]
an equation whose solutions correspond to the eigenfrequencies $\omega_{k}$.
Minor-expanding the determinant using the last column of $\mathbf{M}_{N}$
leads to the recursion formulas
\begin{equation}
m_{n}\equiv\det\mathbf{M}_{n}=-\cos\left(k_{n}x_{n+1}\right)f_{n}+\sin\left(k_{n}x_{n+1}\right)g_{n},
\end{equation}
with
\begin{equation}
\left\{ \begin{array}{ccc}
f_{n} & = & \sin\left(k_{n}x_{n}\right)h_{n-1}+z_{n}\cos\left(k_{n}x_{n}\right)m_{n-1},\\
g_{n} & = & \cos\left(k_{n}x_{n}\right)h_{n-1}-z_{n}\sin\left(k_{n}x_{n}\right)m_{n-1},\\
h_{n} & = & -z_{n}\sin\left(k_{n}x_{n+1}\right)f_{n}-z_{n}\cos\left(k_{n}x_{n+1}\right)g_{n},
\end{array}\right.
\end{equation}
subject to the initial conditions
\[
m_{-1}=\sin\left(k_{-1}x_{0}\right)\qquad\mbox{and}\qquad h_{-1}=-z_{-1}\cos\left(k_{-1}x_{0}\right).
\]
These formulas allow the numerical evaluation of the determinant for
an arbitrary value of $\omega$. For a given geometry, the eigenfrequencies
are numerically determined by first sweeping through the values of
$\omega$, with a certain increment $\delta\omega$, until $\det\mathbf{M}_{N}$
changes sign, bracketing an eigenfrequency whose value is then refined
by the bisection method. The process is repeated with decreasing values
of $\delta\omega$ until no additional eigenfrequencies up to a previously
set value $\omega_{\mathrm{max}}$ are found. For a given eigenfrequency
$\omega_{k}$, the corresponding coefficients $\left\{ A_{jk},B_{jk}\right\} $
(with a new index $k$ to indicate the dependence on $\omega_{k}$)
are determined recursively from Eqs. (\ref{eq:interfacebc}) and (\ref{eq:reflectbc}),
supplemented by the normalization condition $B_{-1,k}=1.$ 

The general solution of the full wave equation then takes the form
\begin{equation}
\Phi\left(x,t\right)=\sum_{k}\phi_{k}X_{k}\left(x\right)\cos\left(\omega_{k}t\right),\label{eq:gensolapp}
\end{equation}
in which 
\begin{equation}
X_{k}\left(x\right)=A_{jk}\cos\left(\omega_{k}x/c_{j}\right)+B_{jk}\sin\left(\omega_{k}x/c_{j}\right),
\end{equation}
with $j$ such that $x_{j}\leq x<x_{j+1}$. The constant coefficients
$\phi_{k}$ are derived from the initial condition $\Phi\left(x,0\right)$
by using the orthogonality condition satisfied by the $X_{k}\left(x\right)$,
\begin{equation}
\sum_{j=-1}^{N}\rho_{j}\int_{x_{j}}^{x_{j+1}}dxX_{k}\left(x\right)X_{q}\left(x\right)=\xi_{k}\delta_{kq}.\label{eq:orthogcond}
\end{equation}
Explicitly, we have
\begin{equation}
\phi_{k}=\frac{1}{\xi_{k}}\sum_{j=-1}^{N}\rho_{j}\int_{x_{j}}^{x_{j+1}}dx\Phi\left(x,0\right)X_{k}\left(x\right),
\end{equation}
with $\xi_{k}>0$ defined by Eq. (\ref{eq:orthogcond}). The conclusion
that the orthogonality condition involves the densities $\rho_{j}$
comes from integrating, over the entire system, the differential equation
satisfied by $X_{k}\left(x\right)$, multiplied by $X_{q}\left(x\right)$,
then exchanging the roles of $q$ and $k$, and subtracting the results,
taking into account the continuity of the pressure at the interfaces.
The above orthogonality condition follows if $\omega_{k}\neq\omega_{q}$.

\bibliographystyle{apsrev}
\bibliography{end}

\end{document}